\documentclass[multphys,vecphys]{promult}

\usepackage{makeidx}     
\usepackage{graphicx}    
\usepackage{multicol}    

\makeindex             

\begin{document}

\title*{Three years of experience with the STELLA robotic observatory}
\titlerunning{Three years of STELLA} 
\author{Thomas Granzer\inst{1}, Michael Weber\inst{1}, \and Klaus G. Strassmeier
\inst{1}}
\institute{Astrophysical Institute Potsdam (AIP), An der Sternwarte 16, D-14482 Potsdam} 
\maketitle

Since May 2006, the two STELLA robotic telescopes at the Iza{\~n}a observatory
in Tenerife, Spain, delivered an almost uninterrupted stream of scientific data.
To achieve such a high level of autonomous operation, the replacement 
of all
troubleshooting skills of a regular observer in software was required. Care must be taken
on error handling issues and on robustness of the algorithms used. In the
current paper, we summarize the approaches we followed in the STELLA
observatory.

\section{Introduction}
\label{sec:1}
STELLA is a fully autonomous robotic observatory with two 1.2m az-alt telescopes 
located at the Iza{\~n}a observatory in Tenerife, Spain (\cite{strassmeier2004},
\cite{strassmeier2006}).
STELLA-I is a classical f/8 Cassegrain-type telescope, equipped with a swiveling M3
mirror to make both Nasmith foci available. STELLA-II is a highly specialized
telescope with the single purpose to feed as much light as possible into an
on-axis fiber. It has a single spherical
mirror at f/1.95. At F1, a field corrector matches the PSF to the fiber entrance
f-ratio and additionally provides a field of view of
roughly 2 arcmin. around the fiber. For acquisition and guiding,
STELLA-II features an auxiliary, 15cm refracting telescope.
STELLA-I currently feeds the STELLA Echelle spectrograph (SES, \cite{weber2008}), but
in late 2009 the new wide-field imaging photometer (WiFSIP) should be
operated on this telescope. STELLA-II is currently in a testing phase.\\
Both telescopes are truly robotic telescopes in the sense that they
autonomously react to changing weather conditions, including operation of the
telescope sheltering building. The object selection is not based on a single,
prescribed sequence, but is always constructed dynamically, making fast reaction
easy (see section
\ref{sec:2}). Internet connection is only necessary for data retrieval and
upload of new targets to the target pool, both of which can be achieved
with considerable low bandwidth. Remote observing is possible, but has so far
never been necessary for normal scientific observations.\\
In the next sections, a step-by-step receipt the how a human observer has been 
replaced
by individual pieces of software is presented. 

\section{Scheduling observations}
\label{sec:2}

Different to classic observatories, robotic observatories in general and STELLA
especially, do not devide the available observing time into different time
slices and dedicate these to single users. All targets are active
at any time, allowing for observing campaigns that span months and, quite
usually, even years. An overview on the different projects conducted at STELLA
might help to mitigate the scheduling requirements.
One of the key projects on the spectroscopic telescope is
the investigation of stellar magnetic activity on a handful of stars.  
The rotational
period defines the time-scale at which a couple of, say, twenty
observations should occur, at intervals as regular as possible. The usability 
of a single observation highly depends on all the other observations occurring
around it: Only if a high phase coverage 
could be achieved, the individual observation was useful. Consequentially, the
scheduling algorithm must be able to adjust the priority of such an observing
project: Once started, it should be assured that it can also finish in time. If
in doubt, refrain from starting the project at all. What comes to ease here is
the general insensitiveness to the starting point of such programs, the only focus
lies on a proper phase coverage. 

Studying highly phase critical phenomena like
the Blazhko effect in RR Lyrae star require the observations to be timed
exactly around certain, well known phases. Here, no freedom in choosing the
starting point is possible, but observations do not need to be clumped
especially close together.

Observations of radial velocity curves of $\delta$ Cephei stars or extra-solar
planets relax that even further. Here you aim at good phase coverage, but the
spacing between individual observations does hardly matter. 

Objects with a prior unknown periods should be scheduled such that no bias is introduced on
subsequent
period determination. 

The final class of targets is those that
introduce no special timing or periodicity, but rather yield highest importance
if observed at optimal conditions -- optimal can refer to certain
astronomic conditions, like no moon-light pollution or minimum airmass 
(easily to predict) or certain seeing requirements (difficult to predict). 

Algorithms that deal with optimized scheduling are not
confined to robotic telescopes alone. Basically all robotic processes face 
similar problems (e.g. \cite{pinedo1995}). Very different approaches to the
scheduling problem exist in literature, those relevant to robotic telescopes
will be discussed briefly.

\subsection{Queue scheduling}

Queue scheduling is the simplest approach possible, but also the least flexible.
A superior process, most likely a human, defines a schedule for the upcoming
observing period (not too long to make reaction to bad weather periods possible,
not too short to gain advantage of the automated observing process). On
observation start, the queue is loaded into the system and followed
task-by-task. A coupling with additional constraints (target must be above a
certain height; target may only be observed within an hour of the scheduled
time) may allow for a limited degree of flexibility, which make queue scheduling
apt for single-task surveillance projects.

\subsection{Critical-path scheduling}

A scheduling algorithm that splits a single task, like an entire observing
campaign, into different, atomic, sub-tasks with various dependencies amongst
each other is known as critical-path scheduling (e.g. \cite{hendrikson1989}).
It can be seen as the mathematical description of a Gantt chart and is mainly
used in huge construction projects, where the main focus lies on the dependency
between individual sub-tasks. Still, the main task of scheduling the different 
task
relative to each other remains and such critical-path scheduling is seldomly 
used in robotic
astronomy.

\subsection{Optimal scheduling}

An optimal schedule describes the particular flow of observations that allows 
the maximization of the scientific return measured according to a predefined
metrics like shutter-open time. The high number of possible permutations of $N$
targets, $p(N)\propto N!$, makes any algorithm a highly demanding
one. In particular, unpredictable changes in environmental conditions 
break the optimal schedule, and a recalculation is necessary. At ground-based
observatories, an
optimal scheduling schema is difficult to implement due to unforeseen 
changes in weather conditions. The 
Hubble Space Telescope, however, uses a software package called SPIKE 
(\cite{giuliano1998}) that delivers an optimal schedule for 14-day periods. 
Attempts to use the
same algorithm on ground-based facilities, e.g. with the Very Large Telescope
(\cite{chavan1998}) or the Subaru telescope (\cite{sasaki2004}) have yielded some 
success in producing guidelines for night 
astronomers.

\subsection{Dispatch scheduling}

The algorithm that schedules targets in real-time, always according to
the current (observational) conditions, is known as dispatch scheduling. 
From the
entire pool of targets available, the algorithm calculates a per-target merit,
picking then the target with the highest yield (Eqn.~\ref{eqn:dispatch}). 
\begin{equation}
m(t)=\sum_i w_i\cdot f_i(t),
\label{eqn:dispatch}
\end{equation}
The summation is done over individual merits $f_i(t)$, with different weights
$w_i$ for a particular merit. A
balanced choice on the individual weights and merits allows for a very
capable scheduling algorithm.
This approach allows easy reaction on changing weather conditions and at the same time 
optimizes target scheduling for distinct side goals. 
However, it has no predictive capabilities in
the sense that the currently top-rated observation will be done, regardless
of an even higher yield possible in the future. 
Nevertheless, this approach is probably suited best for ground-based telescopes as
the reaction to changing weather conditions is algorithm-inherent. Dispatch
scheduling is thus used in many robotic telescope (e.g. \cite{fraser2006}).
On the STELLA observatory it is applied in a somewhat modified approach to
compensate for the bad long-time behavior, see
Eqn.~\ref{eqn:dispatch-stella}. 
\begin{equation}
m(t)=\prod_i v_i\cdot s_i(t) \cdot \sum_j w_j \cdot g_j(t),
\label{eqn:dispatch-stella}
\end{equation}
Here, the $s_i$'s, weighted with constant factors $v_i$, allow long-term modification of target selection 
(i.e. over several nights), while the $g_j$'s are
mainly used for short-term scheduling, i.e. over the course of a given night.
On STELLA, the target itself defines which merits it may use. This is similar
to setting all weights on all non-specified merits to zero, but allows easier
adaptation to new observing strategies: new merits may be added at any time,
given that they are available at run-time. Out of convenience, 
all $s_i$'s and $g_j$'s are limited within $0\leq s_i(t), g_j(t) \leq 1$,
but merits exceeding one are allowed by adjustment of the weights. Generally, three
set of weights for all $s_i$'s and $g_j$'s are available, reflecting the three
principal priority levels: level \emph{A} for high-priority targets, all
observations requested for a single target 
must be completed to allow scientific conclusions. Level \emph{B}
indicates mid-priority targets. Observing strategy is best-effort based,
with (currently) a high likeliness of all observations to be completed. The
lowest priority, \emph{C}, is designed for targets that either add some
scientific value, if observed, or for low-priority targets in large surveys. As
a matter of fact, targets within this priority class are mostly observed during partly clouded
nights, when targets in higher priority bins fail.

From the vast possibilities that the different combinations of weights and merits
allow, only a few possible setups are currently implemented on STELLA. Large
surveys on different targets have a pure airmass merit that peaks at one on the
target's culmination and drops to zero at the apparent horizon, which lies
between $5^\circ$ and $30^\circ$. Depending on the priority class, the weights
are $w_j=1,0.67,0.33$~(A,B,C). Once observed \emph{successfully}, their 
toggle-merit $s_i(t)$ drops to zero. Surveys, where targets should be revisited
after a certain, maybe target-depending, period have an $s_i$ that starts at
one, drops to zero on a successful observation and linearly regains its top value
of one after the specified period.

Targets that need observation strictly at specific phases utilize a so-called
phase merit, that is the sum of normalized Gaussian centered at the requested
phases. Targets that provide no phase zero-point are scheduled according to their
first successful observation. 
Again the weights read as $w_j=1,0.67,0.33$~(A,B,C). In the basic
form, where only strict phasing, but not the evolved time span matters, the
single $s_i$ is constructed such that it starts at $1/N$, where $N$ is the number
of phases requested. As the number $n$ of phases successfully observed increases,
$s_i$ follows $s_i=(1+n)/N$. The accompanying weight $v_i$ equals $v_i=N$, leading to a
gentle increase in total merit from one to $N$ once the target has been started.

\begin{figure}
\centering
\includegraphics[height=6cm,clip]{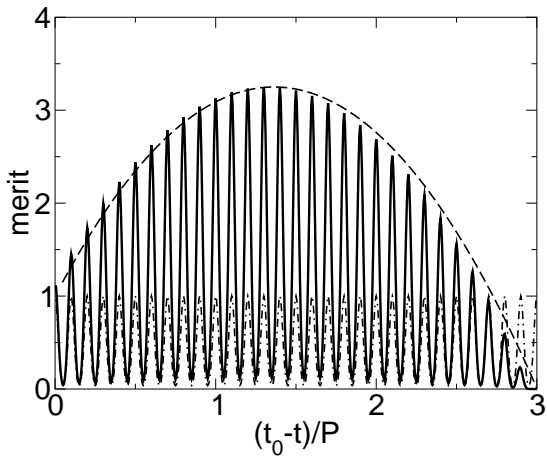}
%
%
\caption{A combination of a long-term increasing and decreasing amplitude
function (dashed line) with a high-frequency phase-selection wave
(dotted-dashed line) to build a selection merit
adequate for scheduling Doppler-imaging targets (thick line).}
\label{fig1}       
\end{figure}

The most ambitious scheduling is done for objects that need a couple of phases
observed within a few periods. At STELLA Doppler-imaging targets, 
these periods are in the range
of several days, leaving you 2-3 weeks to complete a target. The merit is
constructed in the following way. We start with an $s_1$ that peaks around the
opposition of the target, mathematically a normalized Gaussian with a FWHM of
three periods. Once such a target is successfully observed for the first time, 
a second $s_2$ gets activated, which is a parabola fit through points 
$1/v_2$, the
reciprocal of the weight at the starting time and through zero after some multiple of
the period, default after $t=3P$. Additionally, a time until the maximum is
reached can be specified. In addition with the before-mentioned phase-merit,
a complex merit functions like the one depicted in Fig.~\ref{fig1} is reached.
Some more details can be found in \cite{granzer2004}.

%
%

\section{How to carry out a robotic observation}
\label{sec:3}

When a target has been selected by the scheduling process, all the individual
sub-tasks for that particular target have to be carried out -- in parallel
whenever possible, strictly sequential if dependencies between the sub-tasks
exist. Parallelizing many tasks can save valuable observing time, on STELLA,
read-out of the scientific CCD takes place while the telescope already slews to
the next target. In pretty exactly half of the cases, the next target is already
acquired and closed-loop guiding has commenced, when the read-out finishes. 

STELLA houses a spectroscopic and an imaging telescope, thus the individual
sub-tasks for a single observation differ quite substantially, nevertheless the
general idea of splitting an observation into sub-tasks with the possibility to
execute them in parallel or sequentially, stays the same. In STELLA, we
implemented a generic sequencing schema, which is described in detail in
\cite{granzer2004b}. There we also explain how targets can define their own
observing sequence and how a template sequence is constructed. In this paper,
however, we want to focus more on the individual sub-tasks and their implementation, 
particularly pointing out that the solutions have been
chosen for robustness rather than high accuracy.
In the description, we follow the principle time-line of an astronomic
observation, starting by judging the overall weather situation, followed by
pointing and focusing of the telescope, then acquiring of a target, followed by
closed-loop guiding during the scientific exposure. These sections apply
particularly to the fiber-fed instrument Stella-I. Tasks only required for the imaging
telescopes Stella-II follow in the next section,~\ref{sec:4}.

\subsection{Protecting the telescope in harsh weather conditions}

\begin{figure}[htb]
\centering
\includegraphics[height=5cm]{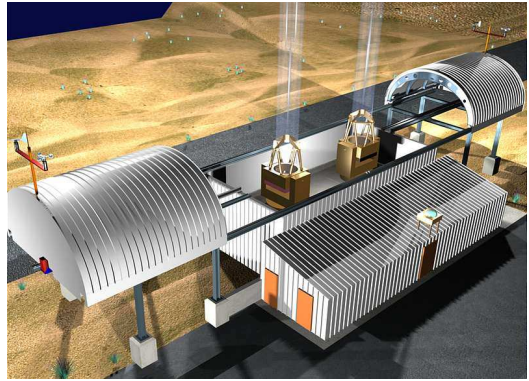}
%
%
\caption{The layout of the STELLA building with the two roll-off roof halves.
The building can be closed independently of the telescopes' positions. This allows
for faster closing times, and, even more importantly, it allows for shutdown also 
in case of telescope movement failures.}
\label{fig2}       
\end{figure}

Protecting the telescope from the outside during bad weather conditions is one
of the task that an autonomous observatory has to fulfill with highest possible
reliability. Needless to say that it is better to lose some observing time at
unstable weather conditions than to risk damage to the telescope or the
instrumentation due to high humidity, rain, high wind, or, particularly
cumbersome in Tenerife, dust. The STELLA observatory is laid out as a
building with a roll-off roof with two, elliptically shaped roof halves driven
by crane motors. Opening and closing of the roof is possible in any position of
the two telescopes, see Fig.~\ref{fig2}. As an additional security mechanism, a 
simple watchdog system that automatically closes the roof in case of
computer crashes is used. 
Systems where the telescope has to be moved to a certain
position before the closing of the roof can commence should be avoided whenever
possible, as it leaves the instruments in an unprotected state if the telescope,
for whatever reason, cannot be turned. 

\begin{figure}[htb]
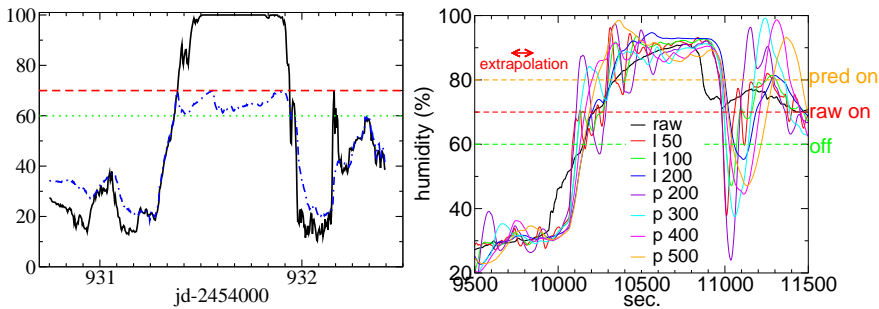

\centering
\includegraphics[height=4cm,clip]{granzer-f3.eps}
\includegraphics[height=4cm,clip]{granzer-f4.eps}
%
%
\caption{Left: The response of the environment system to the outside humidity (thick line). 
When a
level of 70\% on the rising edge is passed, the roofs are closed (note
that the bay humidity, dot-dashed, stays roughly constant thereafter. Re-opening of the
building is allowed, once the humidity has dropped below 60\% and remained at
that level for at least two hours. 
Right: Prediction of inversion layer breakdowns. The thick line is the measured
relative humidity. The 100 second-extrapolations of various extrapolators are super-imposed
onto the true measurement. The true development of the humidity could be foreseen by all of the
extrapolators. Note the delayed onset of the humidity rise of the extrapolated to the true values.
The extrapolation could not predict that a sharp rise will \emph{start} to occur, but
the goal to predict the \emph{height} of the humidity rise is well matched.}
\label{fig3}       
\end{figure}

In standard, i.e. unattended operation mode, the meteorological readings of two
independent weather station decide on the opening or closing of the roof. All
sensor considered critical (precipitation, humidity, temperature, and wind speed) are
available at either station, the redundancy allowing for operation of the
observatory even in case of failure of one of the stations. The building 
control acts completely independent from all other systems and had been
installed even before the telescopes were put in place. In almost
eight years of building operation, it never failed to protect the telescope. 

The decision-making process for the open/close roof process relies on the
current reading of the four critical sensors, as well as on the history of their
measures. A solar-height sensor is a purely calculated sensor, but
enters just like another critical sensor.
Combining two or more sensors in the decision making turned out to
be not necessary.

Most of the critical sensors act as Schmidt-triggers: They toggle their
weather state on two values, depending on the general direction of
the past measures. Toggling
from one state into the other is additionally delayed by a sensor-dependant 
retard time, during that the sensor's reading must stay in the new state,
otherwise toggling does not occur. Typically, this retard time is two seconds or four read cycles
if toggling to the bad weather state -- its main purpose is to
filter out erroneous readings of a sensor. Toggling to the good weather state is more depending 
on the site characteristics. For  Iza{\~n}a, humidity toggles to bad at 70\%, the clear toggle is
set at 60\%. Wind speed toggles at 20m/s and 10m/s, respectively, temperature at -2$^\circ$C and
+1$^\circ$C. Precipitation has just an on/off state, thus zero is considered no rain, one is
considered raining. A brightness sensor toggles at 3000lx and 10000lx. It is mainly a guard
against wrong-posed solar height calculations. 

We apply a retard time of two hours for a humidity event, four hours for
a rain event and twenty minutes for a high wind-gust event. 
This principle is
depicted in Fig.~\ref{fig3}, left, for reaction on the humidity.

Early on, the exposed location of the Tenerife site, just above the inversion
layer, demanded for the capabilities to predict inversion
layer break-downs. For that, the course of the humidity is examined in more detail and
a prognosis for the next 100 seconds -- the closing time of the roofs -- is
derived from it,
see Fig~\ref{fig3}, right panel. Different
time bases and either linear or parabolic extrapolation are used for the near-future humidity 
development. Only if at least six out of
seven extrapolators predict a humidity above 80\% -- compared to the normal toggle of 70\% --, the
roofs close.

\subsection{Pointing the telescope}

\begin{figure}
\centering
\includegraphics[height=5.5cm,clip]{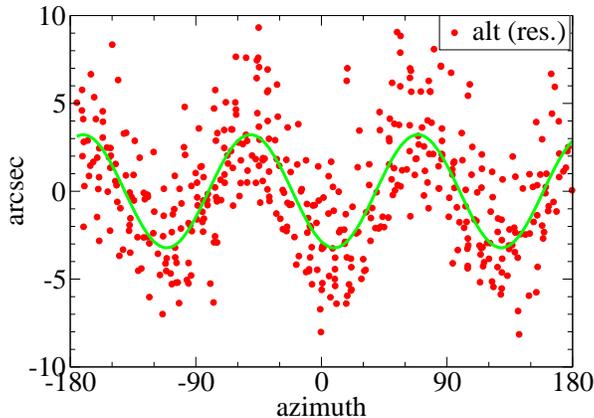}
%
%
\caption{The RMS of the elevation pointings after a classical pointing model has been
applied compared to the true offsets. The RMS varies at an amplitude of $\approx 3$'', at
a wavenumber $k=3$, reminiscent of the tripod mounting of the telescope.}
\label{fig5}       
\end{figure}

For a reliable object acquisition, a good initial pointing of the telescope 
is desirable.
For that, a stable mount is inevitable.
To compensate for the small optical and mechanical misalignments of even the most
precise telescope mounts, a mathematical model known as the pointing model is
used to bring the initial pointing errors down to less than a few arc seconds. STELLA uses
the classical pointing model, which describes only effects of misalignment of an 
otherwise perfect mount. The corrections to the azimuth ($A$) and the
elevation ($E$), namely $\Delta A$ and $\Delta E$, are modeled according to Eqn.~\ref{pm}.
$A_0$ and $E_0$ are constant offsets, the two angles $A_{\rm N}$ and $A_{\rm E}$ describe
the tilt of the telescope's azimuth axis to the true vertical in the northern and eastern direction, respectively.
$N_{\rm PAE}$ describes the non-perpendicularity between the telescope's altitude and azimuth axis.
$B_{\rm NP}$ is the non-perpendicularity between the telescope's altitude axis and the optical axis, while
$T_{\rm F}$ is the tube flexure.
\begin{eqnarray}\label{pm}
\Delta A & = & A_0-B_{\rm NP}\sec E+N_{\rm PAE}\tan E 
         +A_{\rm N}\cos A \tan E+A_{\rm E}\sin A \tan E \nonumber\\
\Delta E & = & E_0-A_{\rm N}\sin A+A_{\rm E}\cos A+T_{\rm F}\cos E
\end{eqnarray}
Ignoring the constant offset $A_0$ and $E_0$, the highest absolute values are for $T_F = 49$'' 
and for 
$B_{\rm NP}$ and $N_{\rm PAE}$ at $B_{\rm NP} = 28$'' and $N_{\rm PAE}=32$'', but note also
the discussion on pointing
models in \cite{strassmeier2007}. One specific effect of the mount of the STELLA
telescopes is depicted in Fig.~\ref{fig5}: the tripod-mooring of the telescope
mount has its resemblance in a 120$^\circ$-wavelength modulated RMS in the 
elevation pointing corrections.
 
\subsection{Focusing with a focus pyramid}

\begin{figure}[tbh]
\centering
\includegraphics[height=4.4cm,clip]{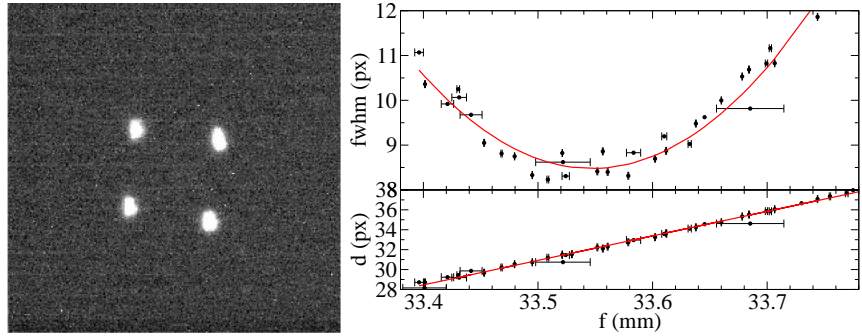}
\includegraphics[height=4.4cm,clip]{granzer-f7.eps}
%
%
\caption{Left: The image of a single star with the focus pyramid introduced into the beam. 
Possible errors
in the center-of-gravity determination of the four image centers do not hamper the 
focussing accuracy due to the high sensitivity of the sub-images'
distances on the focus position.
Right: The calibration curve used to convert the pyramid's sub-image distances to the focus
applied.
The top panel shows the measured FWHM of a star without the pyramid in the optical beam as
a function of focus position (distance of the secondary mirror to M1).
The lower panel shows the average distance of the four sub-images in pixel with the pyramid inside the 
optical beam. This calibration has been done at a very early time in the commissioning of
the telescope, when the telescope control system was still unstable in attaining certain focus
positions. This is visible as the sometimes huge error bars in the focus position.}
\label{fig6}       
\end{figure}

Different solutions for focusing a telescope exists. On Stella-I, we tried to aim at 
a fast and reliable procedure. We decided to equip the telescope with a so-called focus 
pyramid,
which can be rotated into the optical beam. It splits the light of a point-source, i.e. 
of stars,
into four individual sub-images, whose distances are a direct measure of the focus, 
see Fig.~\ref{fig6}.
Once calibrated,  a single measurement of the sub-images' distances suffices to
determine the focus.

\subsection{Acquiring the target}

\begin{figure}
\centering
\includegraphics[height=3.8cm]{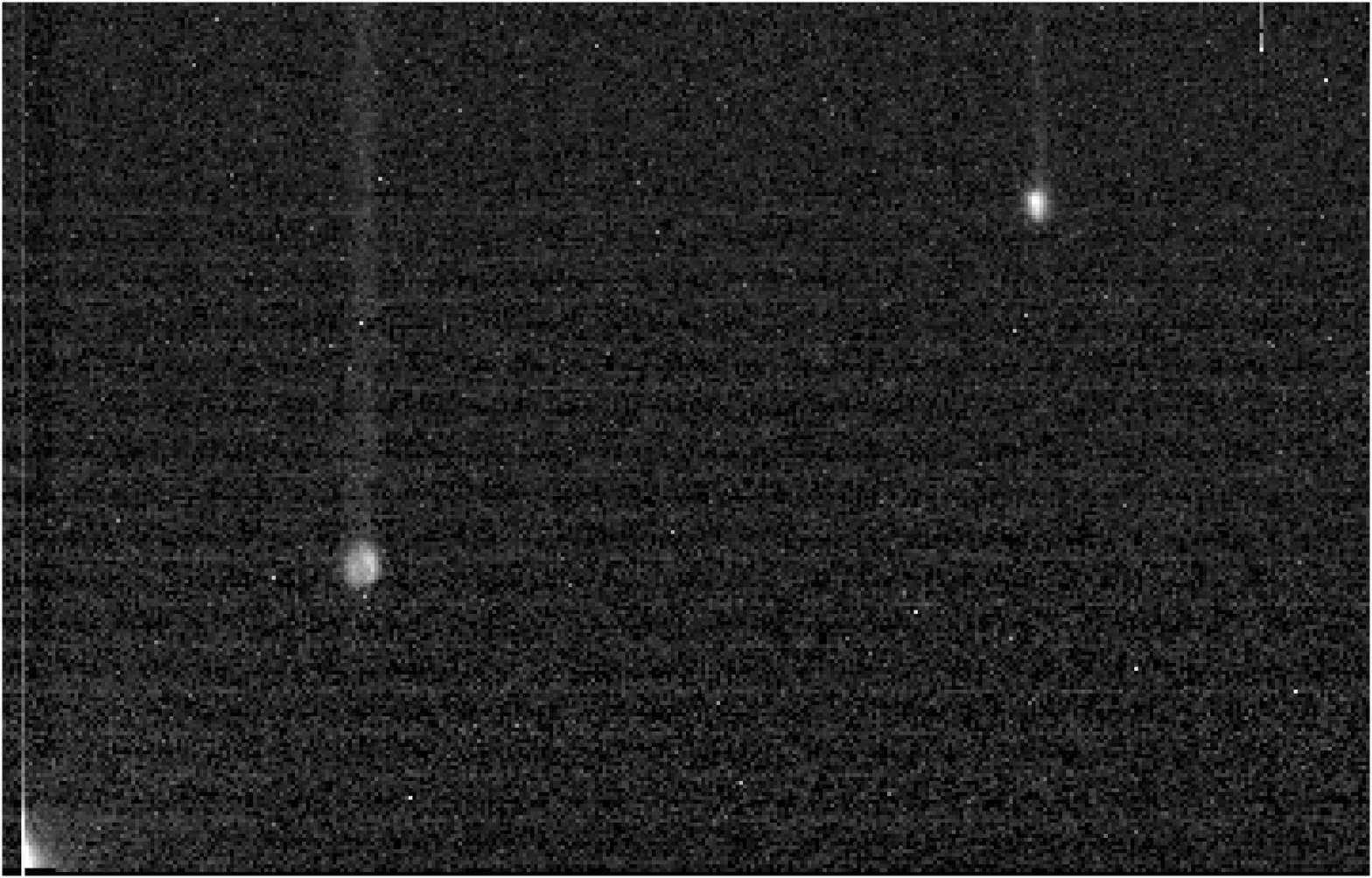}
\includegraphics[height=3.8cm,clip]{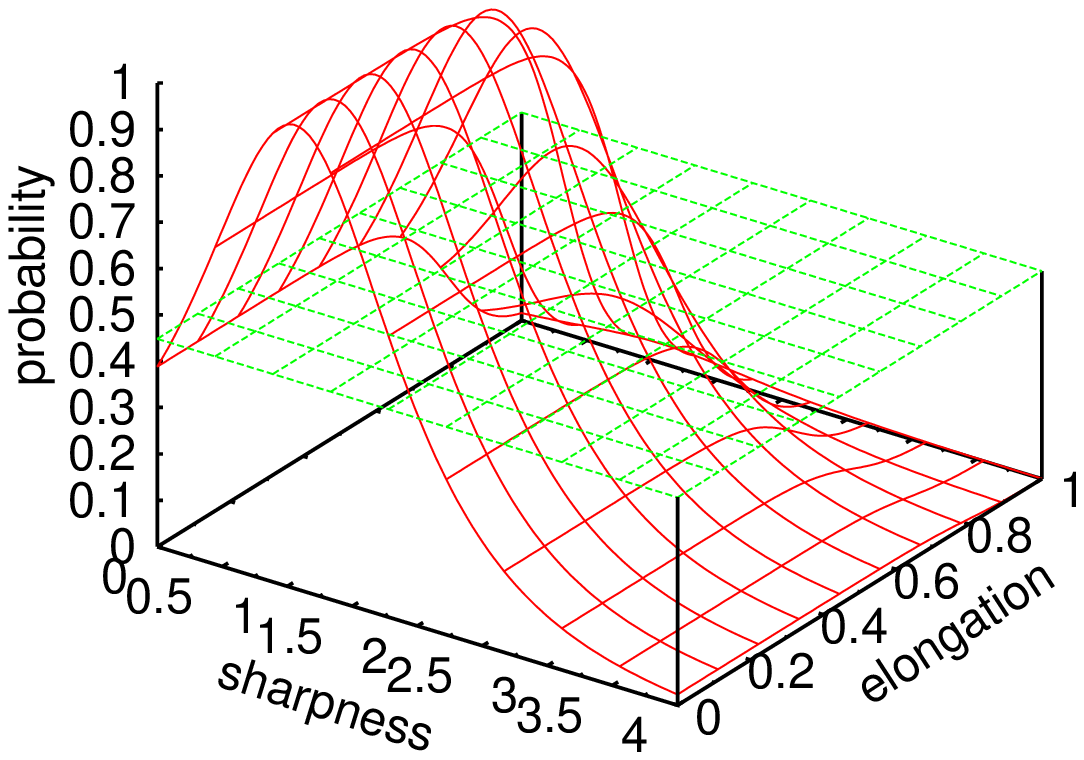}
%
%
\caption{Left:
A typical first image during the acquire phase. The image of the star through the beam 
splitter is the brighter spot to the left, the fainter image to the right is the image from the
mirror around the fiber entrance. The read-out stripes result from the shutter-less design.
Right: The probability function used for identifying stars from their measured image elongation
(y, to the back) and
sharpness (x, to the right).
The green, constant level at a probability of 0.446 is the threshold above which stars are identified. The two-dimensional
function in red was derived by a manual training of the acquire system on 100 different images.}
\label{fig8}       
\end{figure}

Currently, Stella-I is fiber-feeding an Echelle spectrograph 
(\cite{weber2008}).
Thus target acquisition must be done at a precision of less than an arc-second. 
During integration, an adapter unit allows
permanent position control down to sub-arc second levels. This
adapter unit hosts a gray beam splitter that diverts 4\% of the target
light onto a guiding camera. The region just around the fiber entrance is imaged with a mirror 
onto the same guiding camera, leading to a second image at a varying
offset from the direct beam splitter image, see Fig.~\ref{fig8}, left. The guiding camera
has a shutter-less design to minimize the number of moving
parts in the system. The read-out strips inherent to such a design, are dealt
with in the acquisition software.
For details of the optical layout of the acquire unit,
please refer to \cite{strassmeier2004}.

Once the telescope has been moved to the target position, an  image 
of the entire guider's field-of-view (2.1x1.5 arcmin) 
is taken, see Fig.~\ref{fig8}, left. 
The image is bias-subtracted and a truncated Gaussian filter is applied to
it. Similar to DAOfind (\cite{DAO}), the resulting image is used to detect star candidates.
Their sharpness and their elongation is validated against a two-dimensional probability function,
shown in 
Fig.~\ref{fig8}, right. If no stars are found, the initial
acquire is repeated up to five times with gradually increased exposure times. Once an ensemble of
stars has been identified, their positions are matched to the  UCAC2 \cite{UCAC2} catalog and
the shift to the current telescope's position is determined. After the first successful shift,
this procedure is repeated using a 
much smaller window centered around the mirrored fiber entrance to finally reach an offset of
less than 1.5''. This relatively large offset is necessary because image motion induced by
atmospheric turbulence at short exposure times can
amount to an RMS in the target position of up to 0.5''.

\subsection{Closed-loop guiding on the target star}

\begin{figure}
\centering
\includegraphics[width=.95\textwidth]{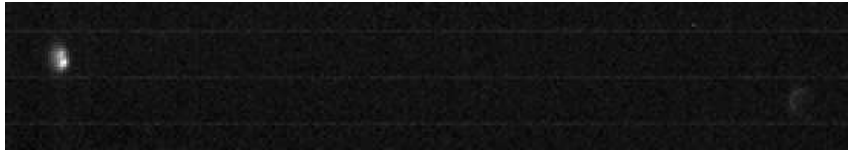}
%
%
\caption{The combined image of $\approx$1200 guiding frames on 51~Peg, total exposure time 20 
minutes.
The bright image to the left is the position of the
star as imaged through the beam-splitter, the fainter image to the right is the light spilled over
the fiber entrance and reflected back onto the guider camera. Guiding is done exclusively
at the brighter image. Aperture photometry of the combined frame is used to  
measure the amount of light lost, here 32\%.}
\label{fig10}       
\end{figure}

Once the star has been successfully acquired, the closed-loop guiding system starts.
Its only aim is to keep the \emph{direct} image of the target star as close to the \emph{mirrored}
fiber entrance
position as possible. The target star's brightness defines the exposure time during the guiding
phase: it is adjusted to get a stellar signal at a S/N ratio of S/N~$\approx 5$. 
The read-out time of the guider window limits the exposure time to 500ms. Typically,
3000 guider frames are taken during a one-hour integration. 
A combined frame of all individual guider frames
can be seen in Fig.~\ref{fig10}. This combined frame is also used to measure the light loss 
at the fiber
entrance by aperture photometry of the two stellar images.

Due to atmospheric image motion, wind shake, and intrinsic telescope oscillations,
not every offset measured should be applied directly to the telescope. In STELLA, a
split approach is used. First, five offsets are averaged. If the average offset is less than 
the standard deviation, or if it is less than the expected image motion induced by the atmosphere,
it is set to zero. This average offset is split into azimuth
and altitude and is fed into two PID controls (from Proportional-Integral-Derivative; for an
introduction to PID refer to e.g. \cite{PID}). 
The output of the PIDs is then applied to the telescope. During commissioning,
three distinct weather situations have been identified, each triggering the use of a different
parameter set.
In normal mode, a proportional term of $P=0.3$ for altitude and $P=0.4$ for azimuth is used.
The integral constant I equals $I=0$, and for the derivative term D, $D_{\rm az} = 0.05$ and
$D_{\rm alt} = 0.02$ is used. The bad-seeing mode, effective when the seeing is worse than 
1.5'',
has a reduced $P_{\rm az} = 0.2$, $P_{\rm alt} = 0.15$ and no D or I term. In high-wind mode for
wind speeds $v > 7$m/s, the number of individual offsets averaged increases from five to 20,
therefore acting on a much slower time-scale. P is thus relatively high, 
$P_{\rm az} = 0.3$, $P_{\rm alt} = 0.25$, and a low I term of $I=0.05$ is introduced. D stays at
zero.

\section{The imaging telescope STELLA-II}
\label{sec:4}

Different to the spectroscopic telescope, 
commissioning on the imaging telescope is just beginning. What follows
are first results and insights gained, but far from being backed by years-long experience as in
the spectroscopic case.
Nevertheless, we want to address a few problems and their possible solutions in high-precision 
robotic photometry.
For a more detailed description on the capabilities of the
Wide Field STELLA Imaging Photometer (WiFSIP), see e.g. \cite{strassmeier-this}.

\subsection{Ensuring high quality flat-fielding}

We do not intend to equip the STELLA building with a flat-fielding screen. The time of twilights 
will be used to obtain sky-flats. The total number of 21 filters available
in WiFSIP makes it impossible to obtain twilight flats on all filters each night. The individual
filter will be grouped together according to their filter system (Sloan, Str{\o}mgren, Johnson-Cousins),
and flats in single, but entire groups should be obtained in a single twilight. 
Two main drivers will define which group of flats
will be chosen, the time passed since the last calibration of the filter
group, and the likeliness of science observations with filters of that group 
in the upcoming night. 

\begin{figure}
\centering
\includegraphics[height=5.5cm,clip]{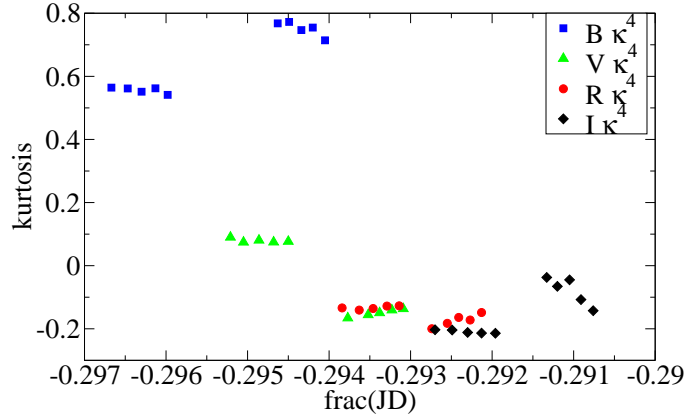}
%
%
\caption{The kurtosis of the ADU statistics versus fractional JD of sky flats taken at the 
Mont Sec observatory
during dusk twilight, five days apart. The different symbols refer to the different Johnson-Cousins 
filters B, V, R, and I. Within each night, the kurtosis stays remarkably constant, allowing for
immediate identifications of flats affected by cosmics or clouds.
The kurtosis offset in the blue filters visible between
the two nights is probably caused by different 
efficiency in the light-straying process in Earths atmosphere.}
\label{fig11}       
\end{figure}

Following the text book, we start with the filter of the shortest central wavelength on dusk
twilight and with the red filter on dawn twilight. Small-window test exposures are taken in rapid
succession until the average ADU level in the read-out window suggests an exposure time in the
allowed range -- 0.2 to 5 seconds. 
The Bonn shutter of the instrument allows for such very short
exposures. On OmegaCam, a similar shutter is reported by \cite{shutter} to deliver equally-exposed
images at exposure times down to 0.1 sec.

To minimize the effect of sky-brightness gradients, the
telescope is pointed to the anti-solar azimuth and five degrees away from the zenith.
This region in the sky is the one with the lowest, sometimes even vanishing
sky brightness gradient, see \cite{chromey1996}. Additionally, the camera will be rotated 
by 180$^\circ$ after half of the individual exposures. Exposure times are gradually changed to
keep the average exposure level as constant as possible.

The quality of the flat-field image is assessed via ADU statistics, with particular
focus on higher-order moments. In Fig.~\ref{fig11}, the kurtosis of the ADUs for five
sky flats, taken in Johnson-Cousins BVRI at two different days are plotted. 
The flats are from the Mont Sec 
robotic observatory (\cite{colome-this}), separated by five days. Though the average 
ADU level dropped from
almost 50000 to below 30000, the kurtosis, or, equivalent the fourth central moment, 
stayed remarkably
constant. Cosmics show up in the kurtosis as high values. Higher moments then the kurtosis work 
even better in detecting cosmics, but also probe the arithmetic
accuracy of the CPU.
 
\subsection{Focusing using statistical moments}

\begin{figure}
\centering
\centerline{\includegraphics[height=3cm]{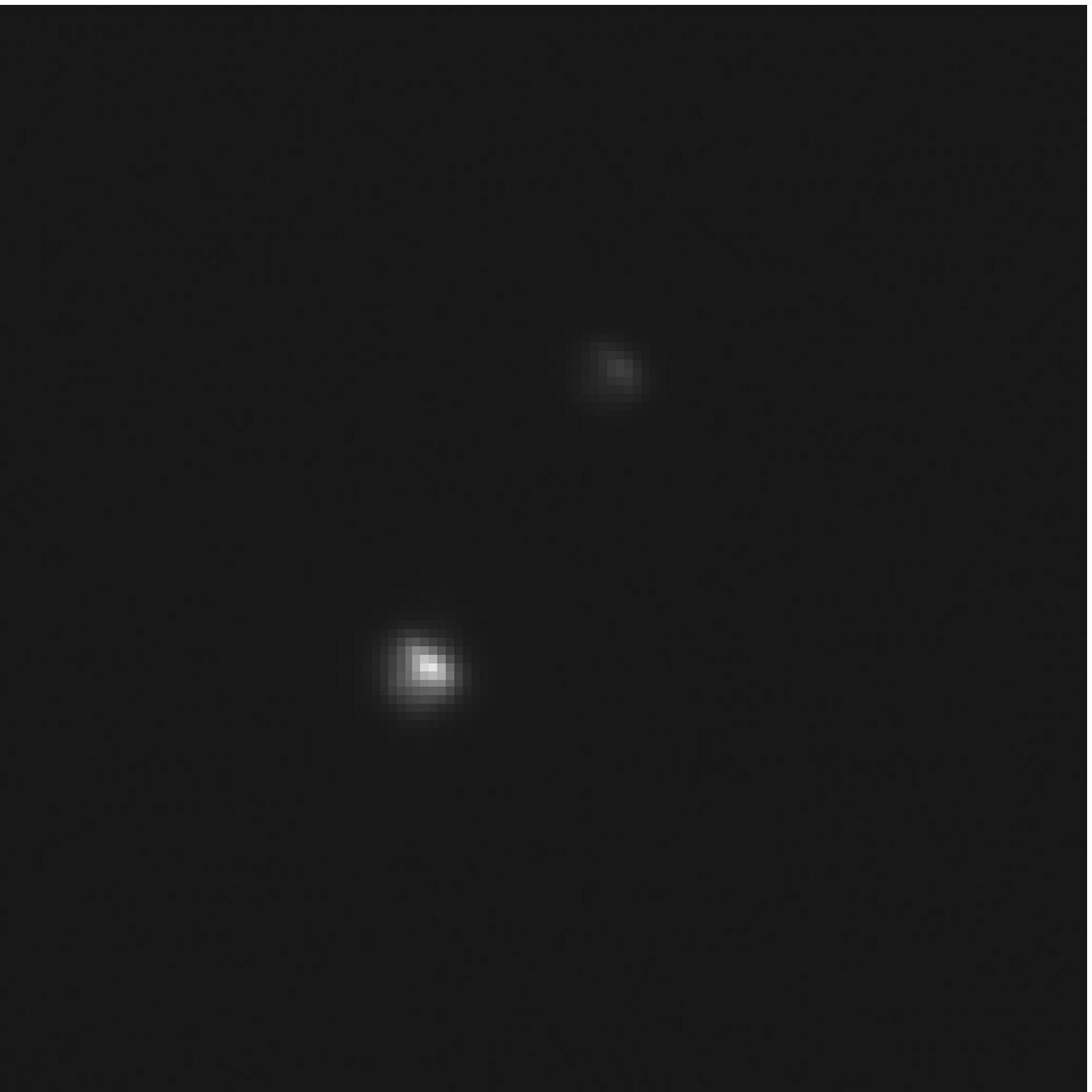}
\includegraphics[height=3cm]{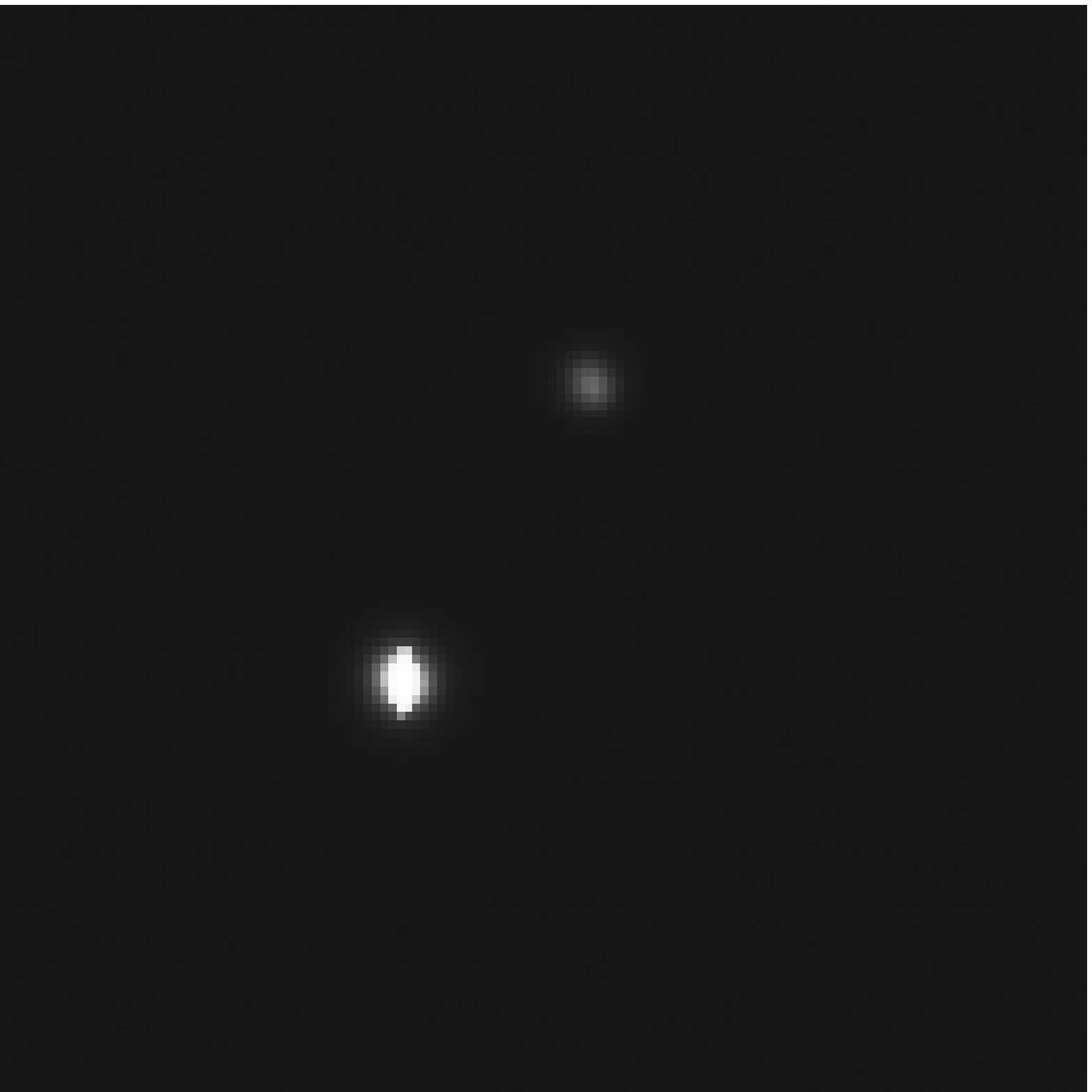}
\includegraphics[height=3cm]{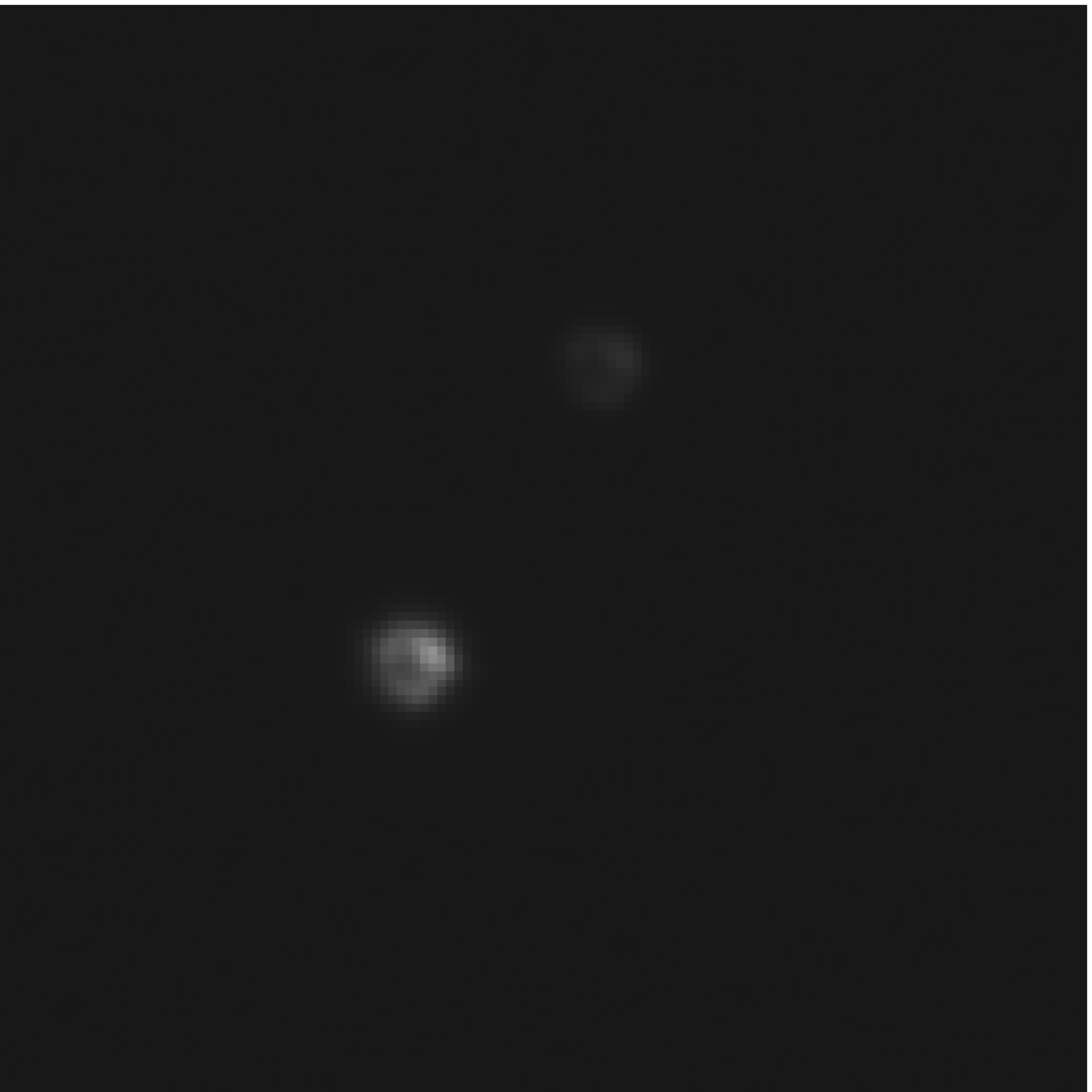}}

\includegraphics[height=6cm,clip]{granzer-f12.eps}
%
%
\caption{The kurtosis of the ADU statistics on three images taken with WiFSIP during
pre-commissioning on Robotel, at the Astrophysical Institute in Potsdam. All three 
quadrants of the imaging CCD with bright stars in them show a peaking kurtosis close
to the best focus position. The one flat line is the kurtosis in quadrant 1, which
was essentially free of stars. The small inserts on top are tiny windows around two bright
stars in the field at the different focus positions.}
\label{fig12}       
\end{figure}

On the imaging telescope, no auxiliary focusing unit like a focus pyramid will
be present. The plan is to use a two-folded approach. Depending on the
temperature of the telescope structure, a focus position will be estimated.
At this estimated focus position together with offsets at $\pm$0.05mm, three images of 
a field close to the celestial North pole (or maybe of Polaris itself, this
will be decided during on-site commissioning) will be taken, and the kurtosis
of the images will be analyzed. Fig~\ref{fig12} shows the anticipated procedure:
The focus will be found at the point maximizing the kurtosis of the image.
We hope that this procedure is superior to directly minimizing the FWHM
of stellar images.

\section{Future plans}
\label{sec:5}

In spring 2010, the spectrograph-feeding fiber will be moved from Stella-I to the second 
STELLA telescope. The light coupling occurs at F1, leaving
no place for a large acquire field. Acquiring will be done using a piggy-back
auxiliary telescope, a 15cm refracting telescope equipped with a shutter-less,
uncooled guiding camera. Again, details on the optical layout can be found in
\cite{strassmeier2006}.

\subsection{Guiding on spilled-over light}

\begin{figure}
\centering
\includegraphics[height=3.7cm,clip]{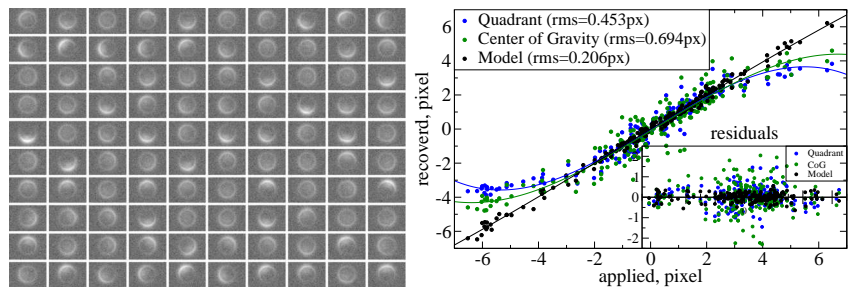}
\includegraphics[height=3.7cm,clip]{granzer-f14.eps}
%
%
\caption{Left: Sequence of artificial images of an R=8$^m$ star on the fiber viewing
Unibrain 520b camera on Stella-II, exposure time set to 1 second. 
Pixel scale is 0.13 arc-sec/pixel, the diameter of the pinhole is 2.8 arc seconds.
The seeing was set to 1 arc-sec.
Right: Recovering of the shift of the center of the star relative to the center of
the pinhole, for the image sequence to the left. Center of gravity uses first
image moments, the quadrant method compares the ADU sum in the left-to-right and up-to-down
image segments. This two methods do not deliver a true shift, but only a shift direction. For
clarity, the direction has been scaled to match the true offset. Direct modeling of the star
plus pinhole recovers the true shift and allows to estimate the light loss and seeing, but 
requires relatively high S/N levels.}
\label{fig13}       
\end{figure}

The main acquiring and guiding system will be built
around the current acquire and guiding logic, but with one additional 
complexity. Due to unavoidable, relative bending of the
auxiliary telescope to the main telescope, guiding with the auxiliary telescope alone will not be
sufficient. Additionally, we will
use the light spilled over on the fiber entrance as a secondary guiding signal. 
This light will
be caught on a fast read-out Unibrain 520b video camera. This
camera comes with an electronic shutter, allowing for exposure times
lower than 1ms and up to 65sec. Furthermore, the gain of the CCD can be
adjusted quasi-continuously from 2.8 to 0.11. This extreme range will
suffice to guide on stars as bright as zeroth magnitude down to,
say, 12$^m$. First tests of the camera on Robotel allowed to estimate the 
quantum efficiency (QE) of the entire system to QE$\approx 0.15$.
Using artificial 
images, see Fig.~\ref{fig13}, we tested three algorithms to recover
the shift of the stellar image with respect to the center of the pinhole:
Center of gravity (CoG, based on image moments), quadrant weights (QW), where the
total ADUs in the left-to-right and up-to-down sectors have been
compared to give a correction direction, and direct modelling of
the star plus pinhole. Only direct modelling returns a true shift,
CoG and QW can only deliver the principal direction of the correction. As
guiding is done in a PID environment anyhow, this shortcoming does hardly
matter as it is compensated by a proper choice of the proportional
term.

The results of all three methods are shown in
Fig.~\ref{fig13}, right panel. Basically
all of the three methods could be used to retrieve the original shift.
The higher robustness favors CoG and QW over direct modelling, while the
latter delivers as a side-product also information on light-loss and
stellar FWHM. Consequentially, we will use CoG during the guiding process
and model the star plus pinhole on the combined frame to extract the additional
information on light loss and seeing.

%
%
%
%
 
%
%
%
\end{document}